\documentclass [12 pt]{article}
\def\be{\begin{equation}}
\def\eea{\end{eqnarray}}
\def\bea{\begin{eqnarray}}
\def\ee{\end{equation}}

\def\la{\langle}
\def\ra{\rangle}
\usepackage[psamsfonts]{amssymb}
\usepackage{amsmath}
\author{F. Kheirandish$^{1}$ \footnote{fardin$_{-}$kh@phys.ui.ac.ir}
\\ $^{1}$ {\small Department of Physics, University of Isfahan,}
\\ {\small Hezar Jarib Ave., Isfahan, Iran.}}
\title{On decaying rate of a quantum state}
\begin{document}
\maketitle
\begin{abstract}
\noindent Decaying rate of a quantum system investigated using the
Fubini-Study definition of distance between states.
\end{abstract}
\section{Introduction}
In Dirac notation a pure state $|\psi(t)\ra$ can be written in an
$N$ complex dimensional vector space \\
\be\label{e1}
 |\psi\ra=\sum_{i=1}^{N}z_i|i\ra,
\ee where $|i\ra$ belongs to a given orthonormal basis and
$z_i$'s are complex numbers.\\
In quantum mechanics, states differ only by an overall complex
factor $\alpha$ are equivalent\\
\be\label{e2} \vec{z}=(z_1,z_2,\cdots,z_n)\sim
\alpha(z_1,z_2,\cdots,z_n), \hspace{1cm}\alpha\in \mathcal{C}.
\ee This recent relation is an equivalence relation on
$\mathcal{C}^N$ and the set of equivalence classes $[\vec{z}]$ is
by definition the projective space $\mathcal{C}P^N$. The complex
numbers in (\ref{e2}) are known as
homogeneous coordinates.\\
there is a natural notion of distance on $ \mathcal{C}P^N $ called
the Fubini-Study[1] distance or metric. The distance $x$ between
two pure states $|\psi_{1}\ra $ and $|\psi_{2}\ra $ according to
Fubini-Study is given by\\
\be\label{e3}
\cos^{2}(x)=\frac{|\la\psi_1|\psi_2\ra|^2}{\la\psi_1|\psi_1\ra\la\psi_2|\psi_2\ra}=
\frac{|\vec{z_1}\cdot\vec{z_2}|^2}{\vec{z_1}\cdot\vec{z_1}\vec{z_2}\cdot\vec{z_2}}.
\ee
where \\
\bea\label{e4}
|\psi_1\ra&=&\sum_{i=1}^{N}z_{1i}|i\ra,\nonumber\\
|\psi_2\ra&=&\sum_{i=1}^{N}z_{2i}|i\ra, \eea and
$\vec{z_1}\equiv(z_{11},z_{12},\cdots,z_{1N})$,
$\vec{z_2}\equiv(z_{21},z_{22},\cdots,z_{2N})\in\mathcal{C}^N$.\\
\section{Decaying rate of a quantum state}
Let $|\psi(0)\ra $ be the initial state of an arbitrary quantum
system descibed by a Hamiltonian $\hat{H}$, using (\ref{e5}), we
want to find the distance between the initial state $|\psi(0)\ra$
and the final stste $|\psi(t)\ra$ of the system during the time
evolution under Hamiltonian $\hat{H}$. For this purpose we find
the infinitesimal form of the equation (\ref{e1}). Let the
normalized state in time $t$ be $|\psi(t)\ra$ then for an
infinitesimal evolution
\bea\label{e5}
|\psi(t+dt)\ra &=&(1-\frac{i}{\hbar}dt\hat{H})|\psi(t)\ra,\nonumber\\
&=&|\psi(t)\ra -\frac{i}{\hbar}dt\hat{H}|\psi(t)\ra, \eea so
\be\label{e6}
\la\psi(t)|\psi(t+dt)\ra=1-\frac{i}{\hbar}dt\la\psi(t)|\hat{H}|\psi(t)\ra.
\ee From (\ref{e5}) and the assumption $\la\psi(t)|\psi(t)\ra=1$,
we have\\
 \bea\label{e7} \la\psi(t+dt)|\psi(t+dt)\ra
&=&(\la\psi(t)|+\frac{i}{\hbar}dt\la\psi(t)|\hat{H})
(|\psi(t)\ra-\frac{i}{\hbar}dt\hat{H}|\psi(t)\ra),\nonumber\\
&=&1+\frac{(dt)^2}{\hbar^2}\la\psi(t)|\hat{H}^2|\psi(t)\ra, \eea
now from (\ref{e6}) and (\ref{e7}) we find the infinitesimal form
of (\ref{e1})
 \be\label{e8}
 dx=\frac{dt}{\hbar}(\triangle\hat{H})_{\psi},
 \ee
 where
 $(\triangle\hat{H})_\psi=\sqrt{\la\psi(t)|\hat{H}^2|\psi(t)\ra-
 (\la\psi(t)|\hat{H}|\psi(t)\ra)^2}$, is the uncertainty of energy in
 state $|\psi(t)\ra$ and is time-independent for a time-independent
 Hamiltonian.
 According to (\ref{e8}), the velocity of decaying $v_d$ of the state $|\psi(t)\ra$, can be
 defined as
 \be\label{e9}
 v_d=\frac{(\triangle\hat{H})_{\psi}}{\hbar}.
 \ee
 The decaying rate of a state $|\psi(t)\ra$, can be defined via survival amplitude at
 time $t$ given by [2]
 \be\label{e10}
 A_t=\la\psi(0)|\psi(t)\ra,
 \ee
  the survival probability is $|A_t|^2$.\\
  Using (\ref{e1}) and it's infinitesimal form (\ref{e8}), it is
  straightforward to show
    \be\label{e11}
 |A_t|=\cos(\int_{0}^{t}\frac{(\triangle\hat{H})_\psi}{\hbar}dt),
 \ee
  when the system is closed, ( time-independent $\hat{H}$), (\ref{e11})
  becomes
 \be\label{e12}
|A_t|=\cos(\frac{t(\triangle\hat{H})_\psi}{\hbar}),
 \ee
 which can be compared with the Mandelstam-Tamm inequality [2]
 \be\label{8}
 |A_t|\geq\cos(\frac{t(\triangle\hat{H})_\psi}{\hbar}).
 \ee
 So during the evolution of a closed system, the survival probability of the state
 $|\psi(t)\ra$, takes it's minimum value in any instant of time.\\
  Decaing rate $w$ of a state $|\psi(t)\ra$, can be defined as
 \be\label{e13}
 w=\frac{d}{dt}(1-|A_t|^2),
 \ee
 so for $|A_t|$ given by (\ref{e11}), we have
 \be\label{e14}
w=\sin(\int_{0}^{t}\frac{2(\triangle\hat{H})_{\psi}}{\hbar}dt)
\frac{(\triangle\hat{H})_{\psi}}{\hbar}, \ee and for a closed
system \be\label{e15}
w=\sin(\frac{2t(\triangle\hat{H})_{\psi}}{\hbar})
\frac{(\triangle\hat{H})_{\psi}}{\hbar}. \ee

\end{document}